\documentclass[a4paper,11pt]{article}
\usepackage{jheppub}
\usepackage[T1]{fontenc}
\usepackage{amsmath}
\usepackage{esint}
\usepackage{babel}
\usepackage{color}
\usepackage{comment}
\usepackage{subfigure}
\usepackage{cancel}
\usepackage{booktabs, siunitx}
\usepackage{enumerate}
\usepackage{multirow}
\usepackage[numbers,sort&compress]{natbib}
\usepackage{hyperref}
\allowdisplaybreaks
\newcommand{\munich}{Max Planck Institute for Physics, Garching bei M\"unchen, Germany}
\newcommand{\liverpool}{Department of Mathematical Sciences, University of Liverpool, Liverpool L69 3BX, 
U.K.
}

\begin{document}

\preprint{MPP-2024-190}

\title{Analytic evaluation of the three-loop 
three-point form factor of $\operatorname{tr}\phi^3$ in $\mathcal{N}=4$ sYM }

\author[a]{Johannes M. Henn,}
\author[a]{Jungwon Lim}
\author[b]{and William J. Torres~Bobadilla}
\affiliation[a]{\munich}
\affiliation[b]{\liverpool}
\emailAdd{henn@mpp.mpg.de}
\emailAdd{wonlim@mpp.mpg.de}
\emailAdd{torres@liverpool.ac.uk}

\abstract{
We compute analytically the three-loop correlation function of the local operator $\text{tr} \, \phi^3$ inserted into three on-shell states, in maximally supersymmetric Yang-Mills theory. The result is expressed in terms of Chen iterated integrals. 
We also present our result using generalised polylogarithms, and evaluate them numerically, finding agreement with a previous numerical result in the literature. We observe that the result depends on fewer kinematic singularities compared to individual Feynman integrals. Furthermore, upon choosing a suitable definition of the finite part, we find that the latter satisfies powerful symbol adjacency relations similar to those previously observed for the $\text{tr} \, \phi^2$ case.}

\maketitle
\graphicspath{{figs/}}

\section{Introduction}

Form factors, i.e. the correlation function of (the Fourier transform of) a local operator, inserted into on-shell states are interesting objects in quantum field theory. For example, the simplest form factors of a current inserted into two on-shell states can be used to study the anomalous magnetic moment of the leptons~\cite{Aoyama:2017uqe,Laporta:2019fmy}, and they are useful in the study of infrared divergences (see \cite{Collins:1989bt} for a review). In particular they have been used to obtain state-of-the-art results for the cusp and collinear anomalous dimension~\cite{Henn:2016wlm,Boels:2017skl,Lee:2019zop,Henn:2019rmi,Henn:2019swt,vonManteuffel:2020vjv,Lee:2021lkc}.

In this paper we focus on form factors in maximally supersymmetric Yang-Mills theory (sYM). The results obtained in such a toy theory can serve as a blueprint for corresponding QCD form factors. In particular, it is interesting to see which simplifications occur when assembling the full form factor from individual Feynman integrals. Moreover, the methods employed here for dealing with the necessary Feynman integrals in dimensional regularisation are general, and will be useful for future QCD studies. Indeed, the same Feynman integrals are relevant for studying processes involving the coupling of a Higgs boson to three gluons, in an effective field theory setup with a large top quark mass.

In sYM, many studies focused on scattering amplitudes on the one hand, and on correlation functions on the other hand, and on relations between them, in particular in the context of a duality with Wilson loops \cite{Alday:2007hr,Drummond:2007aua,Brandhuber:2007yx,Drummond:2007cf,Alday:2010zy,Eden:2010zz}.
From this viewpoint, form factors represent an interesting `hybrid' object of one off-shell composite operator, inserted into a set of on-shell states. 

There is a rich literature on form factors in sYM, going back to reference~\cite{vanNeerven:1985ja}, 
whose authors studied the stress-tensor multiplet inserted into two on-shell states.
Many of the techniques known from scattering amplitudes, and properties known from them carry over to form factors. For example, on-shell techniques may be used to obtain integrands to high loop orders (cf.~\cite{Yang:2019vag} for a review).
The integrand of the two-point form factor mentioned above is known up to the five-loop level~\cite{Yang:2016ear,Boels:2012ew}, and has been integrated up to four loops~\cite{Gehrmann:2011xn,Huber:2019fxe,Lee:2021lkc}.
Although the integrands of the two-point form factors are fascinating, and their integrated expressions contain important information on anomalous dimensions, their  kinematic dependence is fixed entirely by dimensional analysis, as those two-point form factors depend only on a single scale, $s=(p_1+p_2)^2$. Therefore, it is interesting to study multi-point form factors (i.e., a local operator inserted into $n>2$ on-shell states). Already for $n=3$, one then obtains interesting two-variable functions. The same functions are also relevant for processes such as Higgs plus jet production, or vector boson for jet production.

The simplest three-point form factors to consider is the case of the stress-tensor multiplet, represented e.g. by the component $\text{tr} \, \phi^2$, inserted into three on-shell states, as in \cite{Brandhuber:2012vm}. (For related studies of other local operators, and for more external states, cf.~\cite{Brandhuber:2014ica,Brandhuber:2016fni,Brandhuber:2018xzk,Brandhuber:2018kqb,Ahmed:2019nkj}).
The two-loop result of \cite{Brandhuber:2012vm} is already quite interesting. As was remarked in that reference, the result is closely related to a formula for six-particle MHV amplitudes; moreover, the function space, conveniently described by the symbol method \cite{Goncharov:2010jf}, is rather simple: the symbol is built from six symbol letters only. In the QCD literature, that function space is known under the name of two-dimensional harmonic polylogarithms \cite{Gehrmann:2001jv}. 

Recently, interest in these form factors was reignited due to a number of novel structural observations.
Firstly, it was noticed independently in references~\cite{Chicherin:2020umh} and~\cite{Dixon:2020bbt} that the types of two-loop integrals have a hidden structure: their symbol expressions are not arbitrary words in the six alphabet letters, but rather satisfy certain adjacency relations. Those six adjacency relations forbid specific subsequent appearances of letters, thereby reducing the admissible function space. The authors of reference~\cite{Chicherin:2020umh} interpreted these relations in the context of the surprising fact that the six alphabet letters are associated to a B2/C2 cluster algebra. 
Secondly, in the reference \cite{Dixon:2021tdw} 
it was found that there is a relationship between that form factor, and six-particle MHV amplitudes in sYM (see also~\cite{Dixon:2022xqh}). The precise relationship involves a kinematic map as well as an antipodal duality. (An earlier observed two-loop relationship~\cite{Brandhuber:2014ica} between the two quantities does not appear to generalise.)

An interesting and impressive application of these observations is to turn them into bootstrap assumptions for what structures to expect at higher loop orders.
Doing so has made it possible to bootstrap the stress-energy three-point form factor to five~\cite{Dixon:2020bbt}, and all the way up to eight loops~\cite{Dixon:2022rse}. Those bootstrap results are validated by their agreement with recent independent results about the near-collinear limit obtained via integrability~\cite{Sever:2020jjx,Sever:2021nsq,Sever:2021xga,Basso:2023bwv}.
The above results are all the more remarkable, as individual Feynman diagrams do not satisfy some of the assumptions about the symbol alphabet (and adjacency relations) already at three loops~\cite{Henn:2023vbd}. 

The above results strongly motivate us to investigate the $\text{tr}\,\phi^3$  three-point form factor.
Its two-loop expression was computed in~\cite{Brandhuber:2014ica}.
At three loops, the integrand of this form factor was constructed in~\cite{Lin:2021kht,Lin:2021qol}, 
where it was also evaluated numerically at one phase-space point. 
(Four-loop integrands for the form factor of the stress-tensor are available as well~\cite{Lin:2021lqo}.)
However, to date, no analytic results for the three-loop form factor have been reported in the literature, due to the lack of knowledge of the three-loop non-planar Feynman integrals appearing in it.

In this paper we evaluate the necessary Feynman integrals analytically using the method of canonical differential equations~\cite{Henn:2013pwa}, extending previous results for planar Feynman integrals \cite{DiVita:2014pza,Canko:2020gqp,Canko:2021xmn,Henn:2023vbd}, 
and use them to evaluate the form factor.
These missing integrals are a subset of the non-planar three-loop Feynman integrals relevant to Higgs boson decays to three partons. 
That set of integrals has been recently computed in~\cite{Gehrmann:2024tds}.

We begin by checking the expected singular structure of the three-point
$\text{tr}\,\phi^{3}$ form factor up to three loops, by following
the iterative infrared structure of amplitudes in $\mathcal{N}=4$ sYM, from
the Bern-Dixon-Smirnov (BDS) normalisation~\cite{Bern:2005iz,Henn:2011by}.
When constructing the BDS remainder for $\text{tr}\,\phi^{3}$ at three loops, we observe that certain analytical properties (adjacency relations), which are manifest in the two-loop remainder, are no longer satisfied. 
The existence of such normalisation was recently reported in~\cite{Dixon:2024talk}. This normalisation is inspired by the bootstrapped remainder functions of the three-point $\text{tr}\,\phi^{2}$, constructed up to eight loops~\cite{Dixon:2020bbt,Dixon:2022rse}.
In our work, by constructing the form factor from first principles
and ensuring all adjacency relations are satisfied, 
we identify a normalisation that enforces these conditions
that we refer to as the BDS-like normalisation. 

This paper is organised as follows. 
In Sec.~\ref{sec:three-point form factor}, we give an overview on the three-point form factors $\text{tr}\,\phi^2$ and $\text{tr}\,\phi^3$,
by discussing their kinematic configuration, 
the construction of the finite remainders
through the BDS and BDS-like normalisation,
and summarise analytic properties 
of these form factors. 
Then, in Sec.~\ref{sec:2L_phi3}, 
we review the known finite remainders for the form factor $\text{tr}\,\phi^3$ up to two loops.
For the purpose of calculating the three-loop finite remainders, we extend these results to higher orders in the dimensional regulator
$\epsilon$ (up to transcendental weight six functions). 
In Sec.~\ref{sec:3L_phi3}, we present our main result: 
the three-point $\text{tr}\,\phi^3$ form factor to three loops.
We detail the analytic calculation of the three-loop finite remainder, verify it against the only available numerical evaluation, and provide benchmark numerical results along with two- and three-dimensional plots. We also discuss the analytic properties of this form factor, with a particular focus on our chosen normalisation for the BDS-like normalisation.
Lastly, in Sec.~\ref{sec:discussion}, we draw our conclusions 
and discuss directions and open questions.

\par\medskip In the arXiv submission of this paper, we include ancillary files containing detailed information on the computations presented in the following sections. 
We provide the decomposition of the three-loop form factor $\text{tr}\,\phi^3$ in terms of 
masters integrals (\verb"3L_phi3_UT.m"), together with their integral family definition (\verb"3L_families.m") 
and the basis of
relevant master integrals (with up to nine propagators)
(\verb"3L_MIs.m"), and the analytic expressions of the BDS and BDS-like finite remainder functions in terms of generalised  polylogarithms (\verb"phi3_GPL.m") and Chen iterated integrals (\verb"phi3_CII.m") up to three loops.

\section{Three-point form factors in ${\cal N}=4$ super Yang-Mills}
\label{sec:three-point form factor}

In this section, we recall the main features of the 
form factors that describe the interaction of three on-shell states $\Phi$ and a
gauge invariant operator $\mathcal{O}$.
This form factor is defined as the Fourier transformations of the matrix elements of the operator
taken between the vacuum and three on-shell particle states, 
\begin{align}
\mathcal{F}_\mathcal{O}\left(p_1,p_2,p_3;q\right) = \int d^{D}x\,e^{-\imath q\cdot x}\langle\Phi_{1}\Phi_{2}\Phi_{3}|\mathcal{O}\left(x\right)|0\rangle\,.
\end{align}
External states are on-shell $p_i^2=0$ and the operator carries an off-shell  momentum $q^2\ne 0$.
We can define the kinematic invariants, 
\begin{align}
s_{12} &=(p_{1}+p_{2})^{2}\,,\\
s_{13} &=(p_{1}+p_{3})^{2}\,,\\
s_{23} &=(p_{2}+p_{3})^{2}\,.
\end{align}
Due to momentum conservation, $q=p_1+p_2+p_3$,
only three of them are independent,
\begin{align}
q^{2} &= s_{12}+s_{23}+s_{13}\,.
\end{align}
We work in the Euclidean region, 
\begin{align}
q^2<0\,,\qquad  s_{12}<0\,,\qquad  s_{13}<0\,,\qquad  s_{23}<0\,,
\label{eq:unphysical_region}
\end{align}
and consider the dimensionless ratios, 
\begin{align}
& u=\frac{-s_{12}}{-q^{2}}\,, &  & v=\frac{-s_{13}}{-q^{2}}\,, &  & w=\frac{-s_{23}}{-q^{2}}\,,
\label{eq:def_uvw}
\end{align}
satisfying the condition $u+v+w=1$. The Euclidean region in these variables corresponds to, 
$0<u<1$ and $0<v<1-u$ 
(or $0<v<1$ and $0<u<1-v$ ), where $w$ has been expressed in 
terms of $u$ and $v$.

We focus on the half-BPS form factors $\mathcal{O}(x)=\mathcal{O}_k=\text{tr}(\phi^k)$.
For their calculation, we consider
the perturbative expansion in the 't Hooft coupling for the gauge
group $\mathsf{SU(N_{c})}$, 
\begin{align}
g^{2}=\frac{g_{\text{YM}}^{2}N_{c}}{e^{\epsilon\gamma_{E}}(4\pi)^{2-\epsilon}}\,,
\end{align}
 with $\gamma_{E}$ the Euler-Mascheroni constant,
\begin{align}
\mathcal{F}_{\mathcal{O}_{k}} & =\mathcal{F}_{\mathcal{O}_{k}}^{\left(0\right)}\,c_{k}\,
\mathcal{G}_k\,
\delta^{(4)}(q-p_{1}-p_{2}-p_{3})
\,.
\end{align}
Here $c_{k}$ accounts for the colour structures and $\mathcal{F}_{\mathcal{O}_{k}}^{\left(0\right)}$
for the tree-level form factors. For $k=2$ and $k=3$, we have, 
\begin{align}
 & \mathcal{F}_{\mathcal{O}_{2}}^{\left(0\right)}=\frac{\left\langle 12\right\rangle ^{2}}{\left\langle 12\right\rangle \left\langle 23\right\rangle \left\langle 31\right\rangle }\,, &  & c_{2}=\tilde{f}^{a_{1}a_{2}a_{3}}\,,\\
 & \mathcal{F}_{\mathcal{O}_{3}}^{\left(0\right)}=1\,, &  & c_{3}=\tilde{d}^{a_{1}a_{2}a_{3}}\,,
\end{align}
and the perturbative expansion of $\mathcal{G}_{k}$,
\begin{align}
\mathcal{G}_{k} = \left(1+\sum_{\ell=1}^{\infty}g^{2\ell}\,\mathcal{G}_{k}^{(\ell)}\right)\,.    
\label{eq:G_pert}
\end{align}
Notice that it is expected that under exchange (or permutations) of the external
momenta this form factor remains unchanged. This can also be argued
from the indistinguishability of the three external on-shell states.

These operators are part of a half-BPS multiplet of operators and corresponds to 
 a generalisation of the chiral part of the stress-tensor multiplet~\cite{Eden:2011ku,Eden:2011yp}.
The evaluation of this form factor has been studied up to two loops for all $k$, by explicitly computing it from 
first principles~\cite{Brandhuber:2012vm,Brandhuber:2014ica}.
Very recently, the evaluation of $\mathcal{O}_2$ and $\mathcal{O}_3$ has been performed in a fully numerical framework~\cite{Lin:2021kht,Lin:2021qol},
by employing unitarity-based methods and colour-kinematics duality. In parallel, bootstrapping techniques have allowed for the extension of the calculation of $\mathcal{O}_2$ up to eight loops~\cite{Dixon:2020bbt,Dixon:2022rse}.

Since form factors and, in general, multi-loop scattering amplitudes display infrared~(IR) singularities that originate from soft and collinear configurations 
of the loop momenta, it is possible to predict their IR structure by systematically accounting for
universal operators acting on the same amplitudes at lower loop orders~\cite{Catani:1998bh,Becher:2009qa}.
Thus, one defines a finite remainder function, 
\begin{align}
\mathcal{G}_k = \mathcal{G}_k^{\text{BDS}}\exp(\mathcal{R}_k)\,,
\label{eq:bds_exp}
\end{align}
with $\mathcal{G}_k^{\text{BDS}}$ corresponding to the exponentiated one-loop factor,
\begin{align}
\mathcal{G}_{k}^{\text{BDS}}&=\exp\left[\sum_{\ell=1}^{\infty}g^{2\ell}\left(f^{\left(\ell\right)}\left(\epsilon\right)\mathcal{G}_{k}^{\left(1\right)}\left(\ell\epsilon\right)+C^{\left(\ell\right)}\right)\right]\,.
\end{align}
This leads to consider the perturbative expansion of the finite remainder, starting at two loops, 
\begin{align}
\mathcal{R}_k &= \sum_{\ell = 2}^{\infty}g^{2\ell}\, \mathcal{R}^{(\ell)}_k\,.
\end{align}
Explicitly, by combining Eqs.~\eqref{eq:G_pert} and~\eqref{eq:bds_exp},
the two- and three-loop finite remainder functions admit the form, 
\begin{align}
\mathcal{R}^{\left(2\right)}_k&=\mathcal{G}_{k}^{\left(2\right)}\left(\epsilon\right)-\frac{1}{2}\left(\mathcal{G}_{k}^{\left(1\right)}\left(\epsilon\right)\right)^{2}-f^{\left(2\right)}\left(\epsilon\right)\mathcal{G}_{k}^{\left(1\right)}\left(2\epsilon\right)-C^{\left(2\right)}+\mathcal{O}\left(\epsilon\right)\,,
\label{eq:R2}\\
\mathcal{R}^{\left(3\right)}_k&=\mathcal{G}_{k}^{\left(3\right)}\left(\epsilon\right)+\frac{1}{3}\left(\mathcal{G}_{k}^{\left(1\right)}\left(\epsilon\right)\right)^{3}-\mathcal{G}_{k}^{\left(2\right)}\left(\epsilon\right)\mathcal{G}_{k}^{\left(1\right)}\left(\epsilon\right)-f^{\left(3\right)}\left(\epsilon\right)\mathcal{G}_{k}^{\left(1\right)}\left(3\epsilon\right)-C^{\left(3\right)}+\mathcal{O}\left(\epsilon\right)\,,
\label{eq:R3}
\end{align}
with,
\begin{subequations}
\begin{align}
  f^{\left(1\right)}\left(\epsilon\right) & = 1+\mathcal{O}(\epsilon)\,,
  \notag\\
  C^{(1)} & = 0\,,
  \\
  f^{\left(2\right)}\left(\epsilon\right) & =-2\zeta_{2}-2\zeta_{3}\epsilon-2\zeta_{4}\epsilon^{2}
  +\mathcal{O}(\epsilon^3)\,, 
  \\
  C^{\left(2\right)} & =4\zeta_4\,,
  \\
  f^{\left(3\right)}\left(\epsilon\right) &= 4\left[\frac{11}{2}\zeta_{4}+\left(6\zeta_{5}+5\zeta_{2}\zeta_{3}\right)\epsilon+\left(\frac{1909}{48}\zeta_{6}+31\zeta_{3}^{2}\right)\epsilon^{2}\right]
  +\mathcal{O}(\epsilon^3)\,, 
  \\
  C^{(3)} & =16\zeta_{3}^{2}-\frac{181}{3}\zeta_{6}\,.
\end{align}
\label{eq:c_f_values}  
\end{subequations}

Alternatively to the BDS normalisation~\eqref{eq:bds_exp}, 
one can consider a different normalisation (or define a BDS-like 
normalised amplitude), to expose further properties of the 
finite remainder. 
This normalisation was extremely useful to 
expose constraints from the Steinmann relations
in the calculation of six- and seven-point amplitudes~\cite{Caron-Huot:2016owq,Dixon:2016nkn},
and was employed in the computation of the 
leading colour contribution of the form 
factor $\text{tr}\,\phi^2$ through eight loops~\cite{Dixon:2020bbt,Dixon:2022rse}. 
This normalisation can schematically be expressed as, 
\begin{align}
\mathcal{G}_k = \mathcal{G}_k^{\text{BDS-like}}\times\mathcal{E}_k\,,
\end{align}
with,
\begin{align}\label{defBDSlike}
    \mathcal{G}^{\text{BDS-like}}&=\mathcal{G}^{\text{BDS}}\exp\left(-\frac{1}{4}\Gamma_{\text{cusp}}\,\mathcal{E}_{k}^{\left(1\right)}\right)\,,
\end{align}
the cusp anomalous dimension~\cite{Beisert:2006ez}, 
\begin{align}
\Gamma_\text{cusp} = 4g^2 - 8\zeta_2g^4+88\zeta_4g^6
-4\left(
219\zeta_6+8\zeta_3^2
\right)g^8+\mathcal{O}(g^{10})\,.
\end{align}
$\mathcal{E}$ admitting a perturbative expansion 
in the coupling constant, 
\begin{align}
\mathcal{E}_k = \sum_{L=1}^\infty g^{2L}\,\mathcal{E}^{(L)}_k\,.
\end{align}
This amounts to the relation between $\mathcal{E}$ and $\mathcal{R}$,
\begin{align}
\mathcal{E}_k = \exp\left[
\frac{1}{4}\Gamma_\text{cusp}\mathcal{E}^{(1)}_k
+\mathcal{R}_k
\right]\,.
\label{eq:relation_E_R}
\end{align}
In the bootstrapping calculation of the form factor $\text{tr}\,\phi^2$,
it has been observed that the normalised one-loop function $\mathcal{E}^{(1)}$ becomes~\cite{Caron-Huot:2016owq,Dixon:2016nkn,Caron-Huot:2019bsq,Dixon:2020bbt,Dixon:2022rse}, 
\begin{align}
\mathcal{E}_{2}^{\left(1\right)} & =2\left[\text{Li}_{2}\left(1-\frac{1}{u}\right)+\text{Li}_{2}\left(1-\frac{1}{v}\right)+\text{Li}_{2}\left(1-\frac{1}{w}\right)\right]\,.
\end{align}
For $\text{tr}\,\phi^3$, the existence of such normalisation that satisfies similar analytic properties to the form factor $\text{tr} \, \phi^2$ has been reported in~\cite{Dixon:2024talk}.
In this work, by explicitly calculating the three-loop contribution 
to $\mathcal{G}_3$ and ensuring that the analytic properties manifest in  $\mathcal{G}_2$ are also present in $\mathcal{G}_3$, 
we find that the following 
normalisation choice in Eq.~\eqref{defBDSlike} 
leads to the desired properties, 
\begin{align}
\mathcal{E}^{\left(1\right)}_3&=-\frac{1}{2}\left[\log^{2}\left(\frac{u}{v}\right)+\log^{2}\left(\frac{v}{w}\right)+\log^{2}\left(\frac{w}{u}\right)\right]-3\zeta_{2}\,.
\label{eq:E1}
\end{align}
We shall postpone the discussion on the construction of $\mathcal{E}_3^{(1)}$ to Sec.~\ref{sec:analytic_properties},
where we explore in detail the analytic properties we aim to manifest in
$\mathcal{E}_3$,
by taking advantage of the three-loop contribution to $\mathcal{G}_3$.

\section{$\text{tr} \, \phi^3$ three-point form factor to two loops}
\label{sec:2L_phi3}

In this section, we revisit the known results for the form factor
$\text{tr}\,\phi^{3}$ up to two loops, with a focus on the necessary
components for the calculation of the three-loop finite remainder
(see Eq.~\eqref{eq:R3}). The strategies employed in these calculations
are reused in our new three-loop results presented in Sec.~\ref{sec:3L_phi3}.

The analytic expressions of this form factor up to two loops can be cast in the sum of $d\log$ integrals
up to an overall normalisation factor,\footnote{
Here and in the following, black lines correspond to
massless propagators and thick ones to the off-shell
external momenta $q$.}
\begin{subequations}
\begin{align}
\mathcal{G}_{3}^{\left(1\right)} & =\sum_{i=1}^{3}
\mathcal{I}\left(\parbox{20mm}{\includegraphics[scale=0.4]{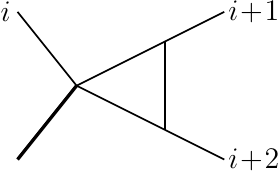}}\, s_{i+1\,i+2}\right)\,,\\
\mathcal{G}_{3}^{\left(2\right)} & =\sum_{i=1}^{3}
\frac{1}{3}\mathcal{I}\left(\parbox{20mm}{\includegraphics[scale=0.4]{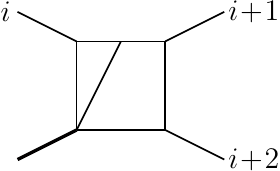}}\,s_{i\,i+1}s_{i+1\,i+2}\right)
+\frac{1}{3}\mathcal{I}\left(\parbox{20mm}{\includegraphics[scale=0.4]{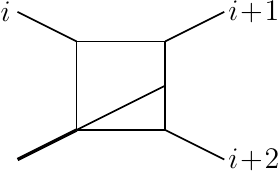}}\,s_{i\,i+1}s_{i+1\,i+2}\right)
\notag\\
&-\mathcal{I}\left(\parbox{20mm}{\includegraphics[scale=0.4]{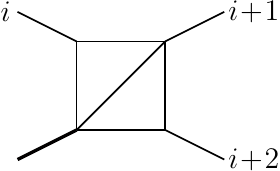}}\,s_{i\,i+2}\right)
+2\mathcal{I}\left(\parbox{20mm}{\includegraphics[scale=0.4]{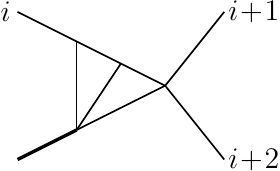}}\,(q^2 - s_{i+1\,i+2})\right)
\notag\\
&-3\mathcal{I}\left(\parbox{20mm}{\includegraphics[scale=0.4]{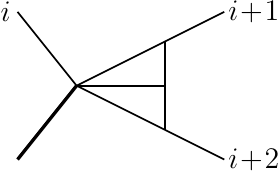}}\,s_{i+1\,i+2}\right)
+\mathcal{I}\left(\parbox{20mm}{\includegraphics[scale=0.4]{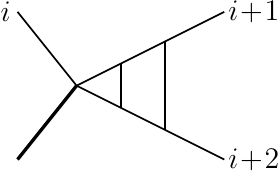}}\,s_{i+1\,i+2}^2\right)\,,
\label{eq:2L_ut_phi3}
\end{align}
\label{eq:1L_2L_ut_phi3}
\end{subequations}
where $\mathcal{I}\left(\mathcal{N}\right)$ corresponds to the Feynman integral 
constructed from the propagators present in the loop topology with numerator $\mathcal{N}$
and $s_{i\,j}\equiv (p_i+p_j)^2$.
Notice that~\eqref{eq:2L_ut_phi3} exactly corresponds to~\cite[Eq.~(3.24)]{Brandhuber:2014ica}
once integration-by-parts identities (IBPs)~\cite{Chetyrkin:1981qh,Laporta:2000dsw} 
are employed to 
express their integrand in terms of our choice of 
master integrals, which by construction 
admit a $d\log$ representation~\cite{Henn:2020lye,Henn:2021aco}. 

The analytic expression of the form factor at one loop is known at all orders in the dimensional regulator $\epsilon$,
\begin{align}
\mathcal{G}_3^{(1)} = S_\epsilon \frac{1}{\epsilon^2}\sum_{i=1}^3\left(-s_{i\,i+1}\right)^{-\epsilon}\,,
\label{eq:1L_G3}
\end{align}
with  $S_\epsilon =e^{\epsilon\gamma_E} \Gamma(1+\epsilon)\Gamma^2(1-\epsilon)/\Gamma(1-2\epsilon)$.
The analogous two-loop expression can be expressed in terms of generalised polylogarithms,
whose set of planar integrals were computed in~\cite{Gehrmann:2000zt} up to transcendental weight four.

Let us now draw our attention to the two-loop contribution to $\mathcal{G}_3$. 
The analytic expression for the finite remainder $\mathcal{R}_3^{(2)}$ has been reported in~\cite{Brandhuber:2014ica},
whose remainder is purely expressed in terms of classical polylogarithms, and their arguments depend on the dimensionless ratios~\eqref{eq:def_uvw}.\footnote{
We remark that in~\cite{Brandhuber:2014ica} the IR subtraction has considered  $C^{(2)}=0$, different to the convention adopted in this paper (see Eq.~\eqref{eq:c_f_values}).
}
Likewise, the normalised two-loop remainder $\mathcal{E}_3^{(2)}$ is obtained 
after expanding~\eqref{eq:relation_E_R} up to order $g^4$, 
\begin{align}
    \mathcal{E}^{(2)} &= \mathcal{R}^{(2)} 
    + \frac{1}{2}\left(\mathcal{E}^{(1)}\right)^2
    - 2\zeta_2 \, \mathcal{E}^{(1)}\,.
\end{align}
This amounts to, 
\begin{align}
\mathcal{E}_{3}^{\left(2\right)}= & \frac{3}{2}\left[\text{Li}_{4}\left(-\frac{uv}{w}\right)+\text{Li}_{4}\left(-\frac{uw}{v}\right)+\text{Li}_{4}\left(-\frac{vw}{u}\right)\right]-3\left[\text{Li}_{4}(u)+\text{Li}_{4}(v)+\text{Li}_{4}(w)\right]\nonumber \\
 & -\frac{3}{2}\log(w)\left[\text{Li}_{3}\left(-\frac{u}{v}\right)+\text{Li}_{3}\left(-\frac{v}{u}\right)\right]\nonumber \\
 & -\frac{3}{2}\log(v)\left[\text{Li}_{3}\left(-\frac{u}{w}\right)+\text{Li}_{3}\left(-\frac{w}{u}\right)\right]\nonumber \\
 & -\frac{3}{2}\log(u)\left[\text{Li}_{3}\left(-\frac{v}{w}\right)+\text{Li}_{3}\left(-\frac{w}{v}\right)\right]\nonumber \\
 & +\frac{9}{16}\left(\log^{4}(u)+\log^{4}(v)+\log^{4}(w)\right)\nonumber \\
 & -\left(\log^{3}(u)\log(vw)+\log^{3}(v)\log(wu)+\log^{3}(w)\log(uv)\right)\nonumber \\
 & +26\left(\log^{2}(u)\log^{2}(v)+\log^{2}(v)\log^{2}(w)+\log^{2}(w)\log^{2}(u)\right)\nonumber \\
 & -\frac{1}{4}\left(\log^{2}(u)\log(v)\log(w)+\log^{2}(v)\log(w)\log(u)+\log^{2}(w)\log(u)\log(v)\right)\nonumber \\
 & +\frac{1}{4}\zeta_{2}\Big[25\left(\log^{2}(u)+\log^{2}(v)+\log^{2}(w)\right)\nonumber \\
 & \qquad-22\left(\log(u)\log(v)+\log(u)\log(w)+\log(v)\log(w)\right)\Big]\nonumber \\
 & +\zeta_{3}\log(uvw)+\frac{199}{8}\zeta_{4}\,.
 \label{eq:E2_logs}
\end{align}
The symbol expansion of the two-loop remainder $\mathcal{E}_3^{(2)}$
becomes, 
\begin{align}
\mathcal{S}\left(\mathcal{E}_{3}^{\left(2\right)}\right)= &
    -\frac{1}{2}u\otimes\frac{u}{(1-u)^{3}}\otimes\frac{v}{w}\otimes\frac{v}{w}-u\otimes v\otimes\frac{u}{w}\otimes_{S}\frac{v}{w}+6u\otimes\frac{u}{v}\otimes\frac{u}{v}\otimes\frac{u}{v}
    \notag\\
    &+\text{perms}\left(u,v,w\right)\,,
\end{align}
in terms of the dimensionless ratios~\eqref{eq:def_uvw},
$\text{perms}(u,v,w)$ accounting for all six permutations of
$\{u,v,w\}$, and the symmetrised tensor product, $x\otimes_{S}y=x\otimes y+y\otimes x$.

\begin{figure}[t]
\centering
\subfigure[]{\includegraphics[scale=0.6]{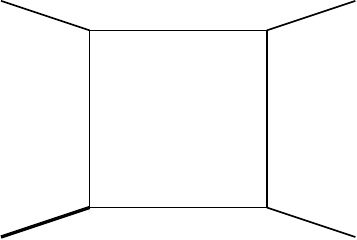}}
\qquad
\subfigure[]{\includegraphics[scale=0.6]{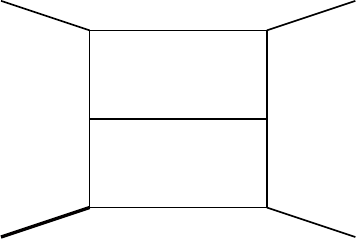}}
\caption{Integral families present in the evaluation of the one- and two-loop form factor $\text{tr}\left(\phi^3\right)$.}
\label{fig:1L_2L_int_fams}
\end{figure}

As a prelude for the construction of the three-loop finite remainder, 
let us remark that one needs to consider
one- and two-loop contributions to $\text{tr}\,\phi^{3}$, according to Eq.~\eqref{eq:R2}. 
While
the one-loop result is known to all orders in $\epsilon$ (see Eq.~\eqref{eq:1L_G3}),
we are required to extend the two-loop contribution to transcendental weight six (or
$\epsilon^{2}$ in the dimensional regulator). The integral families
involved are depicted in Fig.~\ref{fig:1L_2L_int_fams}. 
These integrals are known from references~\cite{Gehrmann:2023etk,Badger:2023xtl}.
In the present work, we independently evaluate these integrals so as to have the results in the same format as the three-loop integrals.
We compute them in the Euclidean region (\ref{eq:unphysical_region}), following the procedure outlined in~\cite{Henn:2023vbd}, and express them in terms of generalised polylogarithms. 
Due to the structure of the form factor, permutations of the external momenta must be considered when calculating these integrals (see (\ref{eq:2L_ut_phi3})). 
The differential equations satisfied by these integrals are solved for the same variables, $u$ and $v$ of Eq.~\eqref{eq:def_uvw}, allowing for straightforward analytic cancellations.

As cross-check of our analytic calculation, 
we recompute the two-loop finite remainders $\mathcal{R}_3^{(2)}$ and 
$\mathcal{E}_3^{(2)}$,
and obtain the three-loop IR structure of 
$\mathcal{G}_3^{(3)}$ from lower loop orders, according to~\eqref{eq:R3}. 
We handle the products of lower loop contributions to $\mathcal{G}_3$
(i.e., $(\mathcal{G}^{(1)}_3)^n$ and 
$\mathcal{G}^{(1)}_3\mathcal{G}^{(2)}_3$
with $n=2,3$) with the aid of {\sc PolyLogTools}~\cite{Duhr:2019tlz},
which are expressed 
in terms of a minimal set of generalised polylogarithms (GPLs)~\cite{Parker:2015cia}. 
We construct this basis up to transcendental weight six functions, 
by systematically enumerating Lyndon words on the sets 
$\{0,1\}$ and $\{0,1,-v,1-v\}$.

With all one- and two-loop contributions at hand, 
we now turn our attention to the complete three-loop calculation 
of the finite remainder, which is the focus of the following section.

\section{$\text{tr} \, \phi^3$ three-point form factor to three loops}
\label{sec:3L_phi3}

In this section, we analytically calculate 
the three-loop form factor $\langle\text{tr}(\phi^3) \phi(p_1) \phi(p_2) \phi(p_3)\rangle$.
We have considered the integrands of~\cite{Lin:2021kht,Lin:2021qol} as our starting point.

The three-loop integrand can pictorially cast as, 
\begin{align}
\mathcal{G}_{3}^{\left(3\right)}  &= 
\mathcal{I}\left(\parbox{25mm}{
\includegraphics[scale=0.4]{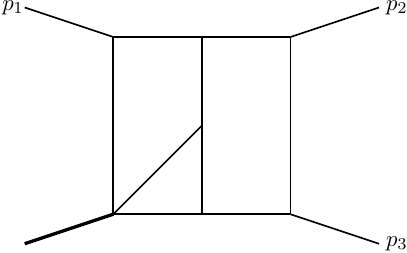}
}\, \mathcal{N}_{2}\right)\,
+ \mathcal{I}\left(\parbox{25mm}{
\includegraphics[scale=0.4]{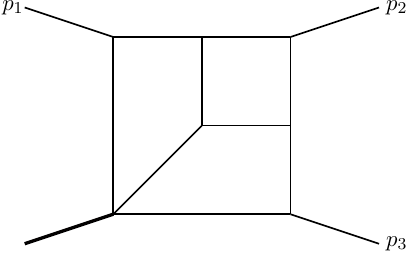}
}\, \mathcal{N}_{3}\right)\,
+ \mathcal{I}\left(\parbox{25mm}{
\includegraphics[scale=0.4]{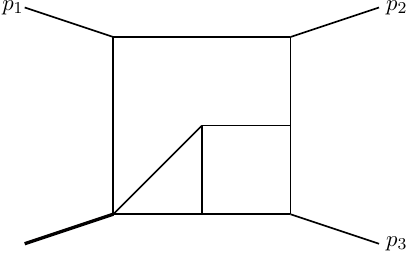}
}\, \mathcal{N}_{4}\right)\,
\notag\\
&+ \mathcal{I}\left(\parbox{25mm}{
\includegraphics[scale=0.4]{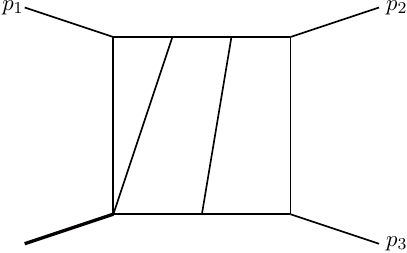}
}\, \mathcal{N}_{5}\right)\,
+ \mathcal{I}\left(\parbox{25mm}{
\includegraphics[scale=0.4]{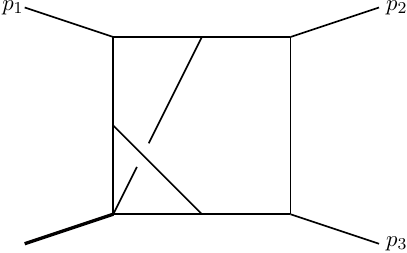}
}\, \mathcal{N}_{9}\right)\,
+ \mathcal{I}\left(\parbox{25mm}{
\includegraphics[scale=0.4]{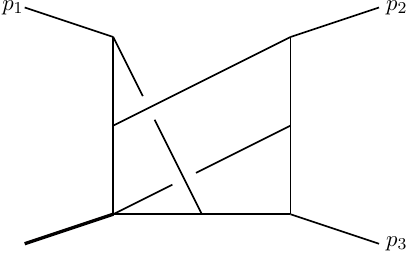}
}\, \mathcal{N}_{11}\right)\,
\notag\\
&+\mathcal{I}\left(\parbox{25mm}{
\includegraphics[scale=0.4]{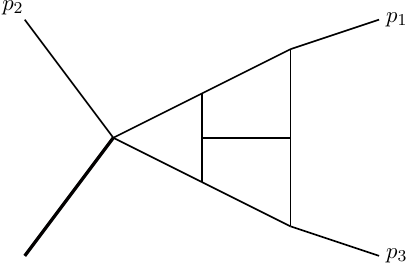}
}\, \mathcal{N}_{21}\right)\,
+\mathcal{I}\left(\parbox{25mm}{
\includegraphics[scale=0.4]{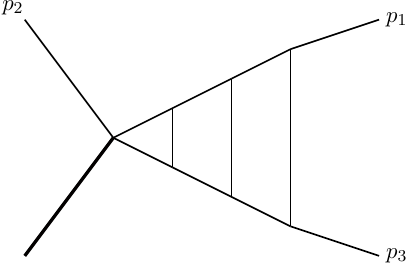}
}\, \mathcal{N}_{22}\right)
+\text{perms}(p_1,p_2,p_3)\,,
\label{eq:3L_phi3}
\end{align}
with ``$\text{perms}\left(p_1,p_2,p_3\right)$''
accounting for all six permutations of the external massless momenta, and $\mathcal{N}$ 
are numerators 
containing dependence on external and internal momenta. 
Here, $\mathcal{N}_i$ exactly matches 
the {\it i}-th diagram of~\cite[Fig.~12]{Lin:2021qol}. 

To analytically evaluate the three-loop contribution 
to $\text{tr}\,\phi^3$, $\mathcal{G}^{(3)}_3$, 
we employ established techniques for calculating multi-loop scattering amplitudes.
We summarise our methodology in Sec.~\ref{sec:analytic_decomposition}. 
We proceed to analytically construct $\mathcal{G}^{(3)}_3$, 
and the remainder functions $\mathcal{R}^{(3)}_3$ and 
$\mathcal{E}^{(3)}_3$, in terms of two-dimensional generalised polylogarithms,
in Sec.~\ref{sec:analytic_evaluation}.
We then compare our results with 
the only available numerical evaluation reported in~\cite{Lin:2021kht,Lin:2021qol}
and present representative two- and three-dimensional plots 
of the finite remainders, in Sec.~\ref{sec:analytic_results}.
In the subsequent Sec.~\ref{sec:analytic_properties}, 
we proceed to analyse their analytic structure to elucidate further properties of 
the $\text{tr} \, \phi^3$ three-point form factor at three loops.
We examine the functional expressions in detail, particularly highlighting the differences between $\mathcal{R}^{(3)}_3$ and $\mathcal{E}^{(3)}_3$, 
and providing a support to the BDS-like normalisation, 
presented in Sec.~\ref{sec:2L_phi3}.

\subsection{Result in terms of master integrals} 
\label{sec:analytic_decomposition}

\begin{figure}[t]
\centering
\subfigure[Integral family A]{\label{fig:famA}\includegraphics[scale=0.65]{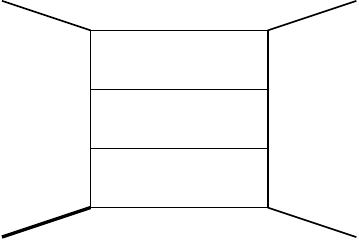}}\qquad
\subfigure[Integral family $E_1$]{\label{fig:famE1}\includegraphics[scale=0.6]{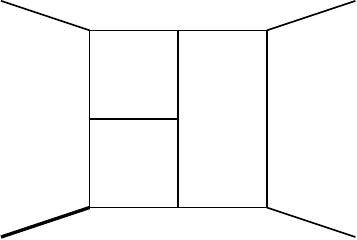}}\qquad
\subfigure[Integral family  $E_2$]{\label{fig:famE2}\includegraphics[scale=0.6]{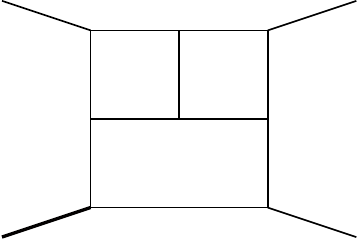}}\\
\subfigure[Integral family  $F_1^*$]{\label{fig:famF1}\includegraphics[scale=0.6]{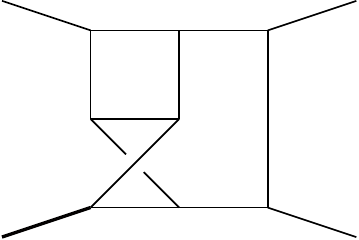}}\qquad
\subfigure[Integral family  $I_2^*$]{\label{fig:famI2}\includegraphics[scale=0.6]{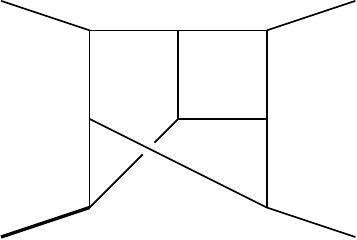}}
\caption{Integral families present in the evaluation of the three-loop form factor $\text{tr}\left(\phi^3\right)$.
For integral families  $F_1^*$ and  $I_2^*$, 
we only consider all possible subtopologies with nine internal propagators. 
}
\label{fig:3L_int_fams}
\end{figure}

\begin{table}[t]
    \centering
    \begin{tabular}{lccccccc}
        \toprule
        \multicolumn{2}{c}{} & 
        \multicolumn{5}{c}{Ordering of $\{p_1,p_2,p_3\}$}\\
        \cmidrule(lr){2-7}        
        {Family}&	{123}&	{132}&	{213}&	{231}&	{312}&	{321}&	{Total} \\
        \midrule
		A &	78 &	73&	46&	71&	44&	40&	352\\
		 $E_1$&	12&	10&	11&	10&	11&	10&	64\\
		 $E_2$&	23&	20&	18&	19&	18&	14&	112\\
		 $F_1^*$ &	32&	26&	28&	22&	23&	19&	150\\
		 $I_2^*$ &	2&	0&	0&	0&	0&	0&	2\\
        \bottomrule
    \end{tabular}
\caption{Decomposition of the three-loop form factor $\text{tr}\,\phi^3$ 
in terms of master integrals according to the ordering of external momenta
in integral family. }\label{tab:3L_mis}
\end{table}

In order to calculate $\mathcal{R}_{3}^{\left(3\right)}$ and 
$\mathcal{E}_{3}^{\left(3\right)}$,
we first evaluate $\mathcal{G}_{3}^{\left(3\right)}$.
We employ the standard approach for computing multi-loop scattering amplitudes. 
We start mapping the integrals appearing in~\eqref{eq:3L_phi3}
onto the families shown in Fig.~\ref{fig:3L_int_fams}, with 
the aid of {\sc Reduze}~\cite{Studerus:2009ye,vonManteuffel:2012np}. 
Then, we apply IBP identities to reduce the integrals to a minimal set of master integrals, chosen to be in canonical form~\cite{Henn:2013pwa}. For this purpose, we employ 
{\sc LiteRed}~\cite{Lee:2012cn} to generate IBPs and sector symmetries between integral families,
and {\sc FiniteFlow}~\cite{Peraro:2019svx} to solve the linear system of equations over finite fields~\cite{Klappert:2019emp,Peraro:2016wsq,vonManteuffel:2014ixa}. 
We find that $\mathcal{G}_{3}^{\left(3\right)}$ is expressed as a linear combination of 680 master integrals, 
which, by construction, are
uniform transcendental degree integrals, and 
whose coefficients are rational numbers.
Thus, displaying that $\mathcal{G}_{3}^{\left(3\right)}$ 
manifests uniform transcendental degree at all orders in $\epsilon$, 
in the very same way as lower loop contributions (see Eqs.~\eqref{eq:1L_2L_ut_phi3}).

In Table~\ref{tab:3L_mis}, 
we summarise the number of master integrals that appear 
in the decomposition of $\mathcal{G}^{(3)}_3$,
according to the order of the external momenta.
An additional explanation of this table is in order. As previously mentioned,
we have implemented sector mappings between all integrals using {\sc LiteRed}.
This is carried out while preserving the order of the integral
families listed in the first column (\{A,  $E_1$,  $E_2$,  $F_1^*$,  $I_2^*$\}) and the
first row (\{\{123\}, \{132\}, \{213\}, \{231\}, \{312\}, \{321\}\}).
Specifically, we map families from the rightmost part of the lists
onto those in the leftmost part of the same list. For example, consider
the integrand with numerator $\mathcal{N}_{9}$ in Eq.~\eqref{eq:3L_phi3} (pictorially corresponding to a nine-propagator subsector of 
the integral family  $F_1^*$), with the external momenta ordered as \{231\}
(i.e. $\{p_{2},p_{3},p_{1}\}$). From this table, it can be seen that subsectors
of  $F_1^*$ (with less than nine propagators)
have been mapped onto A,  $E_1$, and  $E_2$, as well as onto subsectors
that only appear in $F_1^*$ but for different orderings of external momenta, 
such as \{\{123\}, \{132\}, \{213\}\}.

\subsection{Discussion of the Feynman integrals} 
\label{sec:analytic_evaluation}

With the decomposition of $\mathcal{G}^{(3)}_3$ 
in terms of master integrals at our disposal, 
we proceed to analytically evaluate the three-loop Feynman integrals involved in this decomposition (see Fig.~\ref{fig:3L_int_fams}
and Table~\ref{tab:3L_mis}). 
The three-loop planar integrals (A,  $E_1$, and  $E_2$) have already been computed, showing dependence solely on the letters of the alphabet
$\vec{\alpha}_u$~\eqref{eq:uvw_alphabet},
and are expressed in terms of two-dimensional generalised polylogarithms
(cf.~\cite{DiVita:2014pza,Canko:2020gqp,Canko:2021xmn,Henn:2023vbd}).
For the remaining integral families, $F_1^*$ and $I_2^*$, 
we focus on the relevant nine-propagator subsector to obtain integrals with uniform transcendental degree and solve their canonical differential equations for this specific sector. 
A brief comment on the analytic structure of these integrals is in order. While the required subsector of $I_2^*$ depends solely on the letters of the alphabet $\alpha_u$, 
the relevant subsector of  $F_1^*$ exhibits dependence on additional letters
that start appearing at transcendental weight four. 
Remarkably, upon substituting the analytic expressions for the master integrals, we observe pairwise cancellations of these additional letters, resulting in $\mathcal{G}^{(3)}_3$ 
being expressed only in terms of $\vec{\alpha}_u$.
In this paper, we focus on the analytic 
calculation of $\mathcal{G}^{(3)}_3$.
Let us remark that the complete results for all master integrals shown in Fig.~\ref{fig:3L_int_fams}, including those integral families contributing to the three-loop scattering amplitudes relevant to Higgs boson decays to three partons, have been recently computed in~\cite{Gehrmann:2024tds}. For an overview of the framework employed in these calculations, we refer the reader to this thorough reference.

\subsection{Result for remainder function and checks} 
\label{sec:analytic_results}

We obtain the finite remainders $\mathcal{R}^{(3)}$ and $\mathcal{E}^{(3)}$
from the contributions of $\mathcal{G}_3$ up to three loops,
according to Eqs.~\eqref{eq:R3} and~\eqref{eq:relation_E_R}.
Similar to the calculation of the two-loop remainder discussed in Sec.~\ref{sec:2L_phi3}, we observe the analytic cancellation of all infrared poles up to transcendental weight five. 
Thus, we express the finite remainders 
$\mathcal{R}^{(3)}$ and $\mathcal{E}^{(3)}$
in terms of 85 generalised polylogarithms.

\begin{table}[t]
\centering
\begin{small}
\begin{tabular}{cccccccc}
\toprule
$(s_{12},s_{13},s_{23})$ & $\epsilon^{-6}$ & $\epsilon^{-5}$ & $\epsilon^{-4}$ & $\epsilon^{-3}$ & $\epsilon^{-2}$ & $\epsilon^{-1}$ & $\epsilon^{0}$ \\ 
\midrule
$(-2,-2,-2)$ & $\underline{-4.5}$ & $\underline{9.35748}69$ & $\underline{-6.0280}719$ & $\underline{31.503}061$ & $\underline{19.56}3942$ & $\underline{123.5}7978$ & $\underline{216}.98716$ \\ 
$(-1,-1,-2)$ & $-4.5$ & $3.1191623$ & $1.8994028$ & $30.694938$ & $62.552040$ & $189.20903$ & $438.75261$ \\ 
$(-1,-2,-2)$ & $-4.5$ & $6.2383246$ & $-1.3436550$ & $30.266438$ & $40.558842$ & $149.76105$ & $314.26322$ \\ 
$(-1,-2,-3)$ & $-4.5$ & $8.0629176$ & $-4.9111558$ & $32.562137$ & $28.204878$ & $142.41474$ & $262.28451$ \\ 
\bottomrule
\end{tabular}
\end{small}
\caption{Numerical evaluation of $\mathcal{G}_3^{(3)}$ appearing in Eq.~\eqref{eq:3L_phi3}.
The underlined digits show the agreement with the results reported in~\cite[Table~1]{Lin:2021kht}.
}
\label{table:G3_eps_expansion}
\end{table}

\begin{table}[t]
    \centering
    \begin{tabular}{ccc}
        \toprule
        {$(u,v,w)$}&	{$\mathcal{R}^{(3)}$}&	{$\mathcal{E}^{(3)}$} \\
        \midrule
		$(1/3,1/3,1/3)$ & 198.6040626686176 & 53.53642538022867
  \\
		$(1/4,1/4,1/2)$ & 205.3528021206956 & 36.90938731982311 
  \\
		$(1/5,2/5,2/5)$ & 201.0771454928365 & 30.62853770987929
  \\
		$(1/2,1/3,1/6)$ & 205.5332888628428 & 10.44614318690575
  \\
        \bottomrule
    \end{tabular}
\caption{Numerical evaluation of the three-loop finite reminders 
$\mathcal{R}^{(3)}$ and $\mathcal{E}^{(3)}$.
 }\label{tab:3L_num_eval}
\end{table}

In Tables~\ref{table:G3_eps_expansion} and~\ref{tab:3L_num_eval}, 
we present numerical evaluations of 
$\mathcal{G}^{(3)}_3$, and the finite remainders $\mathcal{R}^{(3)}$ and
$\mathcal{E}^{(3)}$ in reference points inside the 
region~\eqref{eq:unphysical_region}.
We find full agreement with the numerical evaluation of the
phase-space point  $(s_{12},s_{13},s_{23})=(-2,-2,-2)$ 
reported in~\cite{Lin:2021kht,Lin:2021qol}.

Let us remark that in~\cite{Brandhuber:2014ica}, 
with the explicit calculation of the two-loop finite remainder $\mathcal{R}^{(2)}$, 
it was observed that this remainder diverges at the boundaries of the region~\eqref{eq:unphysical_region}, 
specifically at $u=0, v=0$ and $u+v=1$. 
These limits correspond to the soft and collinear regimes. 
Below, we revisit these observations for our finite functions
$\mathcal{E}^{(L)}$ at one, two, and three loops.

\begin{figure}[t]
\centering
\includegraphics[scale=0.52]{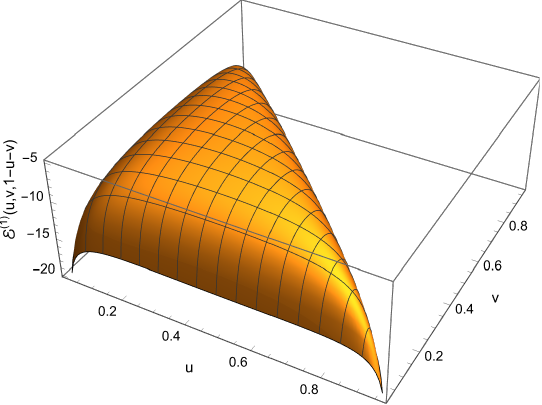}\quad
\includegraphics[scale=0.52]{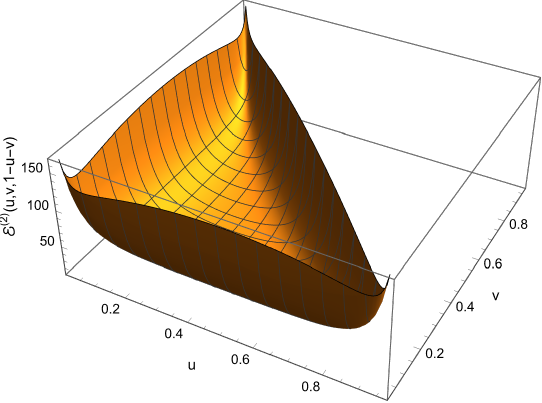}\quad
\includegraphics[scale=0.52]{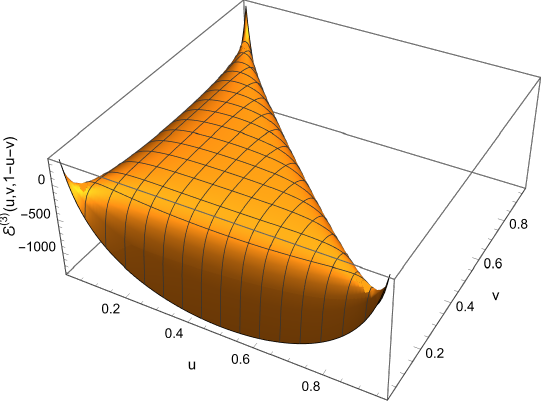}
\caption{Three-dimensional plots of the finite functions 
$\mathcal{E}^{(L)}_3$ at one, two, and three loops  
in region~\eqref{eq:unphysical_region}.
}
\label{fig:3dplots}
\end{figure}

\begin{figure}[h!]
\centering
\includegraphics[scale=0.8]{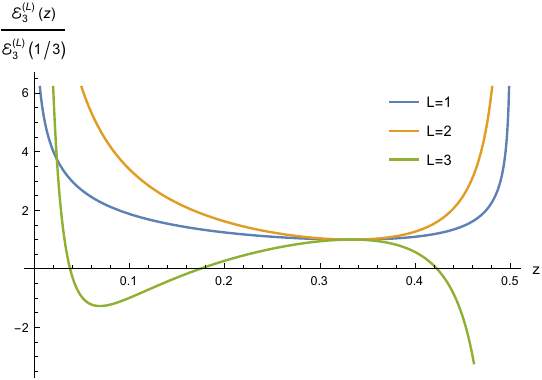}
\caption{Two-dimensional plots of the finite remainder function 
$\mathcal{E}^{(L)}_3$ up to three loops 
in the slice $u=v=z$ with $z\in(0,1/2)$
of region~\eqref{eq:unphysical_region}.
Boundaries on the plot correspond to the soft limit ($z\to 0$) and the collinear limit ($z\to 1/2$), as explained in the text.
}
\label{fig:2dplots}
\end{figure}

In addition to the reference points, we include two- and three-dimensional plots showing numerical stability of our results and illustrating the  behaviour of these finite remainders.\footnote{
We evaluate our analytic expressions with the aid of {\sc Ginac}~\cite{Vollinga:2004sn} through
{\sc PolyLogTools}~\cite{Duhr:2019tlz} and {\sc HandyG}~\cite{Naterop:2019xaf}. 
} 
In Fig.~\ref{fig:3dplots}, 
we plot $\mathcal{E}^{(L)}_3$ up to $L=3$
in the unphysical region ($u,v>0$ and $u+v<1$).
From these plots, we notice that the behaviour when approaching one of the corners, 
say $u=v=z$ (with $z\to0$), 
corresponds to the soft limit of $p_1\to 0$,
\begin{align}
\mathcal{E}_{3}^{\left(1\right)}\left(u,v,w\right) & \overset{u=v=z\to0}{\longrightarrow}-\log^{2}(z)-3\zeta_{2}\,,\\
\mathcal{E}_{3}^{\left(2\right)}\left(u,v,w\right) & \overset{u=v=z\to0}{\longrightarrow}\frac{\log^{4}(z)}{4}+4\zeta_{2}\log^{2}(z)+2\zeta_{3}\log(z)+\frac{77\zeta_{4}}{4}\,,\\
\mathcal{E}_{3}^{\left(3\right)}\left(u,v,w\right) & \overset{u=v=z\to0}{\longrightarrow}\frac{7\log^{6}(z)}{36}-\frac{5}{12}\zeta_{2}\log^{4}(z)-5\zeta_{3}\log^{3}(z)-\frac{157}{4}\zeta_{4}\log^{2}(z)\nonumber \\
 & \qquad\qquad-16\left(\zeta_{2}\zeta_{3}+\zeta_{5}\right)\log(z)+\zeta_{3}^{2}-\frac{3461\zeta_{6}}{48}\,,
\end{align}
whose structure can be appreciated in Fig.~\ref{fig:2dplots},
where we consider the particular slice along the line 
$u=v=z$ (with $0<z<1/2$).

The scenario when approaching a generic point on one of the three edges 
in Fig.~\ref{fig:3dplots} corresponds to a collinear limit. 
For example, $u\to 0$ $(v+w=1)$ is equivalent to $p_1\parallel p_2$,
\begin{align}
\mathcal{E}_{3}^{\left(1\right)}\left(u,v,w\right) & \overset{u=v=z\to1/2}{\longrightarrow}-\log^{2}(2(1-2z))-3\zeta_{2}\,,
\\
\mathcal{E}_{3}^{\left(2\right)}\left(u,v,w\right) & \overset{u=v=z\to1/2}{\longrightarrow}\frac{1}{2}\log^{4}\left(\frac{1}{2}-z\right)+4\log(2)\log^{3}\left(\frac{1}{2}-z\right)\nonumber \\
 & \qquad\qquad+\left(\frac{11}{2}\zeta_{2}+12\log^{2}(2)\right)
 \log^{2}\left(\frac{1}{2}-z\right)
 \nonumber \\
 & \qquad\qquad+\left(22\zeta_{2}\log(2)+\frac{13}{4}\zeta_{3}+16\log^{3}(2)\right)\log\left(\frac{1}{2}-z\right)
 \nonumber \\
 & \qquad\qquad-6\text{Li}_{4}\left(\frac{1}{2}\right)+\frac{31}{4}\log^{4}(2)+\frac{47}{2}\zeta_{2}\log^{2}(2)+\frac{5}{4}\zeta_{3}\log(2)+\frac{89}{4}\zeta_{4}\,,\\
\mathcal{E}_{3}^{\left(3\right)}\left(u,v,w\right) & \overset{u=v=z\to1/2}{\longrightarrow}-\frac{1}{6}\log^{6}\left(\frac{1}{2}-z\right)-2\log(2)\log^5\left(\frac{1}{2}-z\right)\nonumber\\
&\qquad\qquad-2\left(2\zeta_2+5\log^2(2)\right)\log^4\left(\frac{1}{2}-z\right)\nonumber\\
&\qquad\qquad-\left(32\zeta_2\log(2)+\frac{80}{3}\log^3(2)+3\zeta_3\right)\log^3\left(\frac{1}{2}-z\right)\nonumber\\
&\qquad\qquad-\left(\frac{301}{4}\zeta_4+98\zeta_2\log^2(2)+\frac{119}{3}\log^4(2)-8\text{Li}_4\left(\frac{1}{2}\right)+11\log(2)\zeta_3\right)\nonumber\\
&\qquad\qquad\qquad\times\log^2\left(\frac{1}{2}-z\right)+\dots\,,
\end{align}
where the ellipsis corresponds to sub-leading powers of $\log$'s. 
The limit $z\to1/2$ can also be appreciated in Fig.~\ref{fig:2dplots}.

To ensure our paper is self-contained 
and allow the reader to reproduce our results, 
we provide ancillary files in the arXiv submission 
containing the analytic expressions 
of the finite remainders 
$\mathcal{R}^{(2)}_3, \mathcal{R}^{(3)}_3$ and $\mathcal{E}^{(1)}_3, \mathcal{E}^{(2)}_3$,
and $\mathcal{E}^{(3)}_3$. 

In the next section, we discuss the 
analytic properties of the three-loop finite remainder.


\subsection{Analytic properties} 
\label{sec:analytic_properties}
Analytic calculations from bootstrapping approaches 
of the finite remainder $\mathcal{E}_{2}^{(L)}$ 
up to eight loops (with $L=8$),
and the analytic calculation of the two-loop finite remainder
$\mathcal{E}_{3}^{(2)}$
show that these symbol expressions exhibit dependence only on letters of the alphabet, 
\begin{align}
\vec{\alpha}_u=\left\{ u,v,w,1-u,1-v,1-w\right\} \,.
\label{eq:uvw_alphabet}
\end{align}
This alphabet has explicitly appeared in the analytic calculation of four-point two-loop Feynman integrals with 
one off- and three on-shell external states, giving raise to two-dimensional harmonic polylogarithms (2dHPL's)~\cite{Gehrmann:2000zt,Gehrmann:2001ck}.
For the calculation of three-loop Feynman integrals, however, 
it was observed that the alphabet~\eqref{eq:uvw_alphabet} needs to be extended for particular 
integral families~\cite{Henn:2023vbd}. 
Therefore, it is interesting to ask: will higher-loop contributions to the form factor contain letters of the six-letter alphabet~\eqref{eq:uvw_alphabet} only? 
We answer this question by studying 
the analytic properties of the three-loop finite remainder $\mathcal{E}_{3}^{(3)}$. 

With the analytic expression of the 
finite remainder $\mathcal{E}^{(3)}_3$ at hand, 
we can identify the following properties:
\begin{enumerate}[i.]
\item\label{item:1} 
The first entry of the symbol of $\mathcal{E}^{(3)}_3$ always corresponds to an element of the set $\{u,v,w\}$,
which is consistent with physical thresholds for massless processes, $s_{ij}=0$ or $q^2 = 0$. 
\item\label{item:2} 
The last entry of the symbol of $\mathcal{E}^{(3)}_3$ 
corresponds to an element of the set $\{u/v,v/w,w/u\}$.
\item\label{item:3}
The symbol of $\mathcal{E}^{(3)}_3$ is invariant under the action of the dihedral group, generated by the transformations, 
\begin{align}
\text{cycle: }& u\to v\to w\to u\,, \notag\\
\text{flip: }& u\leftrightarrow v\,.
\label{eq:dihedral}
\end{align}
\item\label{item:4}
The symbol of $\mathcal{E}^{(3)}_3$
satisfies a set of six adjacency restrictions, which, in the alphabet $\vec{\alpha}_u$, corresponds to the condition that certain pairs of letters never appear in adjacent positions,
\begin{align}
 & \cancel{\hdots\left(1-u\right)\otimes\left(1-v\right)\hdots}\,, &  & \cancel{\hdots\left(1-v\right)\otimes\left(1-u\right)\hdots}\,,\nonumber \\
 & \cancel{\hdots\left(1-u\right)\otimes\left(1-w\right)\hdots}\,, &  & \cancel{\hdots\left(1-w\right)\otimes\left(1-u\right)\hdots}\,,\nonumber \\
 & \cancel{\hdots\left(1-v\right)\otimes\left(1-w\right)\hdots}\,, &  & \cancel{\hdots\left(1-w\right)\otimes\left(1-v\right)\hdots}\,.
\label{eq:adj_rest_u} 
\end{align}
\end{enumerate}

In order to elucidate further physical constraints
for the symbol of $\mathcal{E}^{(3)}_3$, we use an alternative symbol alphabet, as carried out in the bootstrapping calculation 
of $\text{tr}\,\phi^2$ to eight loops~\cite{Dixon:2021tdw}.
This alphabet is denoted by, 
\begin{align}
\vec{\alpha}_a = \{a,b,c,d,e,f\}\,,
\label{eq:alphabet_a}
\end{align}
with, 
\begin{align}
 & a=\frac{u}{vw}\,, &  & b=\frac{v}{wu}\,, &  & c=\frac{w}{uv}\,,\notag\\
 & d=\frac{1-u}{u}\,, &  & e=\frac{1-v}{v}\,, &  & f=\frac{1-w}{w}\,.
\end{align}
In this representation, the generators of the dihedral group 
are understood as, 
\begin{align}
\text{cycle: } & \left\{ a,b,c,d,e,f\right\} \to\left\{ b,c,a,e,f,d\right\} \,,\notag\\
\text{flip: } & \left\{ a,b,c,d,e,f\right\} \to\left\{ b,a,c,e,d,f\right\}\,,
\label{eq:dihedral_a}
\end{align}
and the adjacency restrictions~\eqref{eq:adj_rest_u}, 
discussed in~\ref{item:4}, correspond to, 
\begin{align}
 & \cancel{\hdots d\otimes e\hdots}\,, &  & \cancel{\hdots e\otimes d\hdots}\,,\nonumber \\
 & \cancel{\hdots d\otimes f\hdots}\,, &  & \cancel{\hdots f\otimes d\hdots}\,,\nonumber \\
 & \cancel{\hdots e\otimes f\hdots}\,, &  & \cancel{\hdots f\otimes e\hdots}\,.
 \label{eq:restr_a_1}
\end{align}
Moreover, this representation of the alphabet enables the identification of additional adjacency restrictions:
\begin{enumerate}[i)]
\setcounter{enumi}{4}
\item \label{item:5}
Specifically, the symbol of $\mathcal{E}_3$ 
never includes the additional set of pairs:
\begin{align}
 & \cancel{\hdots a\otimes d\hdots}\,, &  & \cancel{\hdots d\otimes a\hdots}\,,\nonumber \\
 & \cancel{\hdots b\otimes e\hdots}\,, &  & \cancel{\hdots e\otimes b\hdots}\,,\nonumber \\
 & \cancel{\hdots c\otimes f\hdots}\,, &  & \cancel{\hdots f\otimes c\hdots}\,.
 \label{eq:restr_a_2}
\end{align}
These restrictions follow from the extended Steinmann relations,
which are obeyed by the six-point amplitude in $\mathcal{N}=4$ sYM
theory and supported by the antipodal duality~\cite{Dixon:2021tdw}.
\item \label{item:6}
Lastly, it was found additional restrictions on the triple
sequences of letters that appear in $\mathcal{E}_3$
that are again supported by the antipodal duality:
\begin{align}
 & \cancel{\hdots a\otimes abc\otimes b\hdots}\,, &  & \cancel{\hdots b\otimes cba\otimes a\hdots}\,,\nonumber \\
 & \cancel{\hdots b\otimes bca\otimes c\hdots}\,, &  & \cancel{\hdots c\otimes acb\otimes b\hdots}\,,\nonumber \\
 & \cancel{\hdots c\otimes cab\otimes a\hdots}\,, &  & \cancel{\hdots a\otimes bac\otimes c\hdots}\,.
 \label{eq:restr_a_3}
\end{align}
\end{enumerate}

Let us now draw our attention to the construction
of the normalised function $\mathcal{E}^{(1)}_3$. 
We found that $\mathcal{R}^{(2)}_3$ 
automatically satisfies all properties, indicating that the construction of $\mathcal{E}^{(2)}_3$  should not alter this behaviour. 
This is achieved through the BDS-like normalisation, 
where the one-loop function 
$\mathcal{E}^{(1)}_3$  must satisfy all properties \ref{item:1}--\ref{item:6}, ensuring it does not spoil the properties of $\mathcal{R}^{(2)}_3$.
Accordingly, the possible set of functions for $\mathcal{E}^{(1)}_3$ 
in terms of the letter of the alphabet $\vec{\alpha}_a$ is, 
\begin{align}
    \left\{ \log^{2}\left(a\right),\log^{2}\left(b\right),\log^{2}\left(c\right),\log\left(a\right)\log\left(b\right),\log\left(b\right)\log\left(c\right),\log\left(c\right)\log\left(a\right)\right\}\,.
\label{eq:ansatz_logs}    
\end{align}

The symbol of $\mathcal{R}^{(3)}_3$ automatically satisfies properties \ref{item:1}--\ref{item:4} and~\ref{item:6}, but fails to satisfy property~\ref{item:5}. 
Motivated by the fact of having a remainder function that exhibits all these properties summarised above, we construct an Ansatz based on the functions~\eqref{eq:ansatz_logs}, 
ensuring that the resulting  $\mathcal{E}^{(3)}_3$ satisfies properties \ref{item:1}--\ref{item:6}.
This leads to the functional expression for $\mathcal{E}^{(1)}_3$ presented in Eq.~\eqref{eq:E1}, up to $\pi^2$ terms.

\begin{table}[t]
    \centering
    \begin{tabular}{ccc}
    \toprule
       $L$& $\mathcal{R}_3^{(L)}$ & $\mathcal{E}_3^{(L)}$ \\ 
    \midrule
    1 & -- & \ref{item:1}--\ref{item:6} \\
    2   &  \ref{item:1}--\ref{item:6} & \ref{item:1}--\ref{item:6}\\
    3  & \ref{item:1}--\ref{item:4},~\ref{item:6} & \ref{item:1}--\ref{item:6}\\
    \bottomrule
    \end{tabular}
    \caption{Analytic properties of the finite remainders of the 
    three-point form factor $\text{tr}\,\phi^3$ up to three loops.}
    \label{tab:properties}
\end{table}

Furthermore, we examine properties \ref{item:1}--\ref{item:6} beyond the symbol level. 
This is achieved by expressing the finite remainder $\mathcal{R}_{3}^{(3)}$ 
in terms of Chen iterated integrals~\cite{Chen:1977oja},
which explicitly depend on the letters of the alphabet $\vec{\alpha}_{u}$
(or $\vec{\alpha}_{a}$). 
These integrals of weight $k$ are defined as follows, 
\begin{align}
\left[\alpha_{i_{1}},\hdots,\alpha_{i_{k}}\right]_{\vec{s}_{0}}\left(\vec{s}\right) & =\int_{\gamma}d\log\alpha_{i_{k}}\left(\vec{s}'\right)\left[\alpha_{i_{1}},\hdots,\alpha_{i_{k-1}}\right]_{\vec{s}_{0}}\left(\vec{s}\right)\,,
\end{align}
 with $\left[\right]_{\vec{s}_{0}}=1$, and $\gamma$ a path connecting
the boundary point, $\vec{s}_{0}=\left(u_{0},v_{0}\right)=\left(0,0\right)$,
and another point, $\vec{s}=\left(u,v\right)$.
We observe that once $\mathcal{E}_3^{(3)}$ is constructed 
from the BDS-like normalisation, all properties are satisfied as summarised in Table~\ref{tab:properties}.

\section{Discussion and outlook}
\label{sec:discussion}

In this paper we calculated analytically the three-loop three-point form factor of $\text{tr}\, \phi^3$ in sYM theory. The result is a pure weight six function, expressed in terms of Chen iterated integrals. We found that while the individual Feynman integrals contributing to this calculation depend on more integration kernels, the form factor depends on six integration kernels, or alphabet letters, only. 
Moreover, we found that by choosing a particular definition of the finite part, as anticipated in~\cite{Dixon:2024talk}, 
the latter satisfies a total of twelve adjacency relations~\cite{Dixon:2021tdw}.
(Similar simplifications are conjectured to hold for $\text{tr}\, \phi^2$ form factors in sYM.)

By choosing a particular integration path in the Chen iterated integrals, our result can straightforwardly be expressed in terms of generalised polylogarithms. Evaluating the latter numerically we found perfect agreement with previous numerical results for this form factor.

There are a number of promising directions.
The techniques we have used are applicable also to other three-point form factors, as well as to related  scattering amplitudes involving one massive and three massless external states, especially 
now that the full set of master integrals has become available~\cite{Gehrmann:2024tds}. 
For example, it would be interesting to evaluate form factors of the Konishi operator in sYM, as the latter is related via supersymmetry to the ${\text{tr}}\, F^3$ operator that in QCD arised from the effective coupling of a Higgs boson to three gluons. 
Furthermore, it would be interesting to consider the $\text{tr}\, \phi^2$ operator, which at three loops has both a leading colour as well as a subleading colour component. The leading colour piece was predicted by bootstrap methods~\cite{Dixon:2020bbt,Dixon:2022rse}, 
with expressive checks via consistency as well as integrability. 
A first-principle Feynman diagrams calculation would prove this result. 
The subleading-colour contributions are not known, 
and it would be very interesting to see whether the simple symbol alphabet, and the adjacency relations hold there as well.

\acknowledgments

We would like to dedicate this work to the memory of Stefano Catani, whose inspiring work on the physical structure of scattering amplitudes deeply influenced our thinking. His passion for science and generous spirit will be greatly missed, but his insights continue to shape our work.

We thank Benjamin Basso, Lance Dixon and Alexander Tumanov for correspondence and for sharing their bootstrap results with us prior to publication~\cite{Basso:2024xxx}.
We thank Marcus Spradlin for correspondence, Thomas Gehrmann, Petr  Jakubčík, Cesare Carlo Mella, Nikolaos Syrrakos, and Lorenzo Tancredi for collaboration on closely related projects, and Simone Zoia and Gang Yang for useful discussions.
This research received funding from the Excellence Cluster ORIGINS funded by the Deutsche Forschungsgemeinschaft (DFG, German Research Foundation) under Germany's Excellence Strategy - EXC-2094-390783311, 
the Galileo Galilei Institute for Theoretical Physics (GGI) during the Workshop {\it Theory Challenges in the Precision Era of the Large Hadron Collider},
and by the Leverhulme Trust, LIP-2021-01.

\bibliographystyle{JHEP}
\bibliography{refs}

\end{document}